\documentstyle[11pt,epsfig]{article}\def\ifPNAS{n}

\def\figtest{y}
\def\figsin{y}	% y= figures in; n= omit figs

\long\def\cut#1{{\sf[#1]}}
\long\def\cut#1{}
\long\def\CUT#1{}

\def\Asym{A}	% bend stiffness/kt
\def\Atsym{A\tot}
\def\Psym{P}	% structural persistence
\def\kasym{\kappa_{\rm bend}}%bend stiffness
\newcommand{\rrangle}{\rangle\!\rangle}\newcommand{\llangle}{\langle\!\langle}

\def\lk{{\sl Lk}}

\def\spin{_{{\rm spin}}}\def\rigid{_{{\rm rigid}}}
\def\drag{_{{\rm drag}}}

\def\tot{_{{\rm tot}}}
\newcommand{\inv}{^{\raise.15ex\hbox{${\scriptscriptstyle -}$}\kern-.05em 1}}
\newcommand{\half}{{\textstyle{1\over2}}} %puts a small half in a displayed eqn

\def\crit{_{\rm crit}}
\def\kbt{k_{\rm B}T}
\def\ie{{\it i.e.}}
\def\via{{\em via}}\def\eg{{\em e.g.}}
\newcommand{\eem}[1]{\cdot10^{-#1}}
\def\UV{{\rm erg\,sec\,cm^{-3}}}   %viscosity
\def\UT{{\rm dyn\,cm}}             %torque
\def\figtest{y}
\def\tfig#1{Fig.~\ref{#1}}
\def\ifigure#1#2#3#4{\ifx\figsin\figtest
\begin{figure}[t]\begin{center}{\epsfysize=#4truein \epsfbox{#3}}
\end{center}\smallskip
\caption{{\footnotesize #2}\label{#1}}\end{figure}
\else
\begin{figure}[h!]
\caption{\footnotesize #2\label{#1}}\end{figure}
\fi}
\newcommand{\pitemize}{\begin{itemize}\setlength{\itemsep}{0pt}\setlength{\parsep}{0pt}}
\newcommand{\penumerate}{\begin{enumerate}\setlength{\itemsep}{0pt}\setlength{\parsep}{0pt}}

\ifx\ifPNAS\figtest
\renewcommand{\ell}{l}

\else

\fi

\begin{document}
\author{Philip Nelson\\Department of Physics and Astronomy\\University
of Pennsylvania\\Philadelphia PA 19104 USA\\
{\tt www.physics.upenn.edu/$\sim$nelson/}\\
\ifx\ifPNAS\figtest phone 215-898-7001; fax
215-898-2010\\ e-mail {\tt nelson@physics.upenn.edu}\fi
}
\title{Transport of Torsional Stress in DNA}
\date{30 June 1999; {\bf revised} 18 Oct 1999}
\maketitle

\ifx\ifPNAS\figtest
\bigskip\begin{center}Classification: Biophysics\\
%Text pages: 28; Figures: one; Tables: none\\
%Character Count: $<43190$\\
	\end{center}
\newpage\fi

\begin{abstract}
It is well known that transcription can induce torsional stress in
DNA, affecting the activity of nearby genes or even inducing
structural transitions in the DNA duplex. It has long been assumed
that the generation of significant torsional stress requires the DNA
to be anchored, forming a limited topological domain, since otherwise
it would spin almost freely about its axis. Previous estimates of the
rotational drag have, however, neglected the role of small natural bends
in the helix backbone. We show how these bends can increase the drag
{\it several thousandfold} relative to prior estimates,
allowing significant torsional stress even in linear, unanchored
DNA. The model helps explain several puzzling experimental
results on structural transitions induced by transcription of DNA.

\end{abstract}
\ifx\ifPNAS\figtest\newpage\fi

\section{Introduction and Summary\label{sintro}}
DNA can be regarded as a linear repository of
sequence information, or as a chemical compound subject to
various modifications
(\eg{} methylation), and each of these viewpoints is important for
understanding some aspects of gene function and regulation.
However, many other important processes require an appreciation
of DNA as a {\sl physical elastic object in a viscous environment.}
For example, the action-at-a-distance
between eukaryotic promoters and their enhancers involves an
effective concentration of bound enhancer units depending on both
torsional and bend rigidity of DNA.

While the equilibrium statistical mechanics of stiff
macromolecules such as DNA is a classical topic (see
\eg{} \cite{yamakawabook}), still the {\it nonequilibrium transport}
properties of such molecules remains incomplete, in part due to
the experimental difficulty of probing those
properties. In particular, Liu and Wang proposed that the {\sl
transport of torsional stress (torque)} along DNA during
transcription could play a  role in gene
regulation (the ``twin supercoiled domain model'')
\cite{liu87a}. Transcription causes axial rotation of the
transcribed
DNA relative to the transcribing polymerase. If free rotation is
hindered in some way, a resulting
torsional stress will propagate down the DNA,
destabilizing (or overstabilizing) the double helix structure at
some distant point. The resulting ``topological coupling'' between
nearby genes has been observed in several experiments (see
Sect.~\ref{sexperiments} below).

Liu and Wang assumed a simple mechanism for the transport of
torsional stress, following Levinthal and Crane\footnote{Levinthal and
Crane's ``speedometer-cable'' motion will be called ``plumber's
snake'' motion, or ``spinning'' motion, in the present paper.}
\cite{levi56a}.  In a viscous
medium a straight infinite rod meets a
frictional resistance to axial rotation given by
\begin{equation}
\tau=\mu\spin\omega L
\label{naive friction}\end{equation}
Here the torque $\tau$ (with dimensions of energy) depends on the
rotation rate $\omega$ (radians/sec) and length $L$ \via{} a
friction constant $\mu\spin$. A simple calculation \cite{lamb45a} gives
$\mu\spin=4\pi \eta R^2\approx1.3\eem{15}\,$dyn$\cdot$sec, where
$R\approx1\,$nm is the rod radius
and $\eta=0.01\,\UV$ is the viscosity of water. Other authors give slightly
different prefactors
\cite{garc81a}.

Liu and Wang pointed out that the torsional
friction constant $\mu\spin$ appearing in eqn.~(\ref{naive friction})
is extremely small due
to the factor of $R^2$, and so they concluded that no significant
torsional stress was possible in  DNA of reasonable length without
some additional physical anchoring.
Absent such anchoring, both
linear (open)  and circular (plasmid) DNA would {\it spin in place,}
like a plumber's snake \cite{levi56a}.
For concreteness we will consider below the example of a linear DNA of
length 3.5~kbp (1200~nm),
rotated at its end with angular frequency
$\omega=60\,$radians/sec; a  related case is a
7~kbp construct, linear or circular, rotated near its center. In
either case formula~(\ref{naive friction}) gives a maximum torsional stress
$\tau\approx9\eem{18}\,\UT$. Since the torque needed to denature
DNA locally is several thousand times greater (see below), Liu and Wang's
conclusion
seems to be safe.

The analysis of this paper was motivated by several experimental
observations which defy the familiar analysis just
summarized (Sect.~\ref{sexperiments} below). A variety of assays,
both in living cells and {\it in vitro,}
have found significant torsional stress following
transcription at a single promoter on unanchored DNA
constructs. All these experiments are sensitive to topoisomerase,
pointing to the role of torsional stress.  The estimates given
above imply that such large stresses are impossible.

To resolve this paradox, the analysis in Sects.~\ref{spp}--\ref{scalcs} below
will show that
the classical formula (\ref{naive friction}) can be very misleading:
it {\sl vastly underestimates the torsional stress on the DNA
duplex near the transcribing polymerase.} The discussion rests on the
observation that DNA is a heteropolymer, {\it i.e.} it is naturally
bent on length scales longer than its persistence length of about
50~nm. For a curved molecule to spin in place without dragging
sideways through the surrounding medium, as assumed in formula
(\ref{naive friction}), requires constant {\it flexing}. The natural
bends resist this flexing, forcing the molecule to translate through
the fluid and greatly increasing the viscous drag through the
surrounding water. (\tfig{ffour}{\it d} below summarizes the model.)
This enhanced drag indeed
explains the large observed torsional stress near the point of
transcription.

\section{Experiments\label{sexperiments}}
\subsection{General\label{ssgeneral}}
This section briefly reviews a few of the relevant experimental
results, focusing on {\it in vitro} assays. Sect.~\ref{spp} below
describes our physical model.

RNA polymerases are efficient motors: for example,
{\it E.~Coli} RNAP can
generate forces of up to 20~pN against an opposing load
\cite{wang98a}. When the same mechanical energy is expended against a
{\it torsional} load, it corresponds to a torque of
20~pN$\cdot$0.34~nm/step divided by 2$\pi$ radians for every 10.5~steps,
or $10^{-13}\,$\UT, more than enough to induce structural transitions
in DNA. The speed of transcription ranges from 50~nt/sec in eukaryotes
to twice as great for T7
\cite{iked87a}. The corresponding rotational driving rates are
then $\omega= 30$ and 60 radians/sec, respectively.

The actual torsional stress during transcription need not, however,
attain the maximal value just given. Liu and Wang's twin supercoiled domain
model rests on the observation that torsional
stress will only build up if {\it a)}~the
polymerase itself is prevented from counterrotating about the DNA
template, and {\it b)}~a suitable {\it torsional load} opposes the
rotation of the DNA at a point sufficiently close to the cranking
polymerase. The present paper is concerned mainly with point {\it
(b)}, but for completeness we first digress to discuss {\it (a)}.

\paragraph{Anchoring at the transcribing polymerase}
A number of effects can prevent counterrotation of the polymerase. For
example, in eukaryotes the polymerase may be physically attached to
the nuclear matrix. Even without a rigid attachment, the eukaryotic
polymerase holoenzyme is physically quite large and thus offers a
large hydrodynamic drag to rotation. Similarly in prokaryotes, the
nascent RNA transcript can begin translation before it is fully
transcribed, leading effectively to a large complex consisting of
polymerase, transcript and ribosome. Liu and Wang proposed a
particularly attractive possibility: if the emerging protein is
membrane-bound (for example, the tetracycline-resistance {\it tet} gene
product), it can anchor its ribosome to the cell membrane
\cite{liu87a}. Many experiments have shown that translation of {\it
tet} greatly increases twin-supercoil domain effects (see \cite[and
references therein]{lill96a}).

The above mechanisms operate only {\it in vivo}. Remarkably,
twin-supercoiled domain effects have also been observed in a number of
{\it in vitro} assays, where no cellular machinery exists
(see Sect.~\ref{ssresults} below). At least three mechanisms can
nevertheless create significant drag opposing counterrotation of the
polymerase: {\it i)}~polymerase has been found to create a tight loop
in the DNA, greatly increasing its effective hydrodynamic radius and
hence the drag for counterrotation \cite{tenh92a}; {\it ii)}~the
nascent RNA transcript itself will create some hydrodynamic drag to
rotation \cite{liu87a}; {\it iii)}~under the conditions of most
experiments ({\it e.g.} \cite{tsao89a}) polymerase is present at
concentrations leading to
batteries of simultaneously-transcribing complexes. To relieve
torsional stress, all active
complexes would have to  counterrotate  simultaneously, with a drag
proportional
to their total number.

\paragraph{Anchoring elsewhere}
Thus, even {\it in vitro,} transcription can effectively lead to the
cranking of DNA by a nearly immobilized polymerase. As mentioned in
point {\it (b)}~above, however, cranking at one point  still does not
suffice to create torsional stress:  DNA rotation must be effectively
hindered somewhere else as well, since otherwise both
linear and circular DNA would simply spin freely in place at the
driving rate $\omega$.

As in point {\it (a),} many mechanisms can anchor DNA in the crowded
cellular environment. For example,  in eukaryotes a
DNA-binding protein could tie the DNA onto some part of the nuclear
matrix.
Another possibility, envisioned by Liu and Wang and implemented in
several experiments, is to bind a second polymerase to the DNA and
rely on its resistance to rotation as in {\it (a)}~above. The
second polymerase can either be stalled or actively transcribing in
the opposite (divergent) sense from the first.

Once again, however, the clearest results come from the {\it in vitro}
assays mentioned earlier, in which only a single promoter is active on
a circular \cite{tsao89a,drog91a,drog93a,drol94a,wang97b} or even
linear (L. B. Rothman-Denes, unpublished results; D. Levens, unpublished
results)
template. In these experiments the only known hindrance to free
spinning motion is the torsional hydrodynamic drag. If DNA were
effectively a simple, straight, rod of diameter 2~nm, then the estimate in
equation
(\ref{naive friction}) would apply, and we could confidently predict
that transcription would generate negligible torsional stress. Since the
experiments contradict this expectation, we must modify the na\"\i ve physical
picture of the transport of torsional stress in DNA.

\subsection{Experimental results\label{ssresults}}
\paragraph{In vitro}
Tsao {\it et al.} made a circular plasmid with only one promoter
actively transcribing \cite{tsao89a}. They assayed transient torsional stress
in the
wake of polymerase by
allowing topoisomerase~I to selectively eliminate negative supercoils,
then measuring the remaining degree of positive supercoiling via 2d
electorphoresis.
They found that transcription induces a degree
of supercoiling ``much bigger than expected'' and concluded that ``it is
possible that the
	degree of supercoiling generated by  transcription is
	underestimated in the theoretical
	calculation'' of \cite{liu87a}.

Dr\"oge and Nordheim assayed torsional stress in a 3~kbp circular plasmid using
the B-Z structural transition \cite{drog91a}. They concluded that
``Interestingly our results suggest that diffusion rate of
transcription-induced superhelical twists must be relatively slow
compared to their generation, and that under {\it in vitro} conditions
localized transient supercoiling can reach unexpectedly high levels.''
Similarly Dr\"oge later found that transcription can induce
site-specific recombination {\it in vitro} \cite{drog93a}. Here the
conclusion is that transcription created local torsional stress,
in turn driving local writhing and bringing recombination sites into
synapsis. Wang and Dr\"oge later extended these experiments and called
attention to the fact that torsional strain remains localized in a
gradient region close to the polymerase, instead of spreading rapidly
around the plasmid and cancelling at the antipodal point \cite{wang97b}.

Drolet, Bi, and Liu studied the reciprocal effects of topoisomerase~I
and gyrase \cite{drol94a}, assaying with 1d electrophoresis. The
result of interest to the present paper is that they found that
membrane anchoring via the nascent TetA protein was not necessary for
transcription-induced supercoiling, in contrast to earlier {\it in
vivo} studies.

Finally,
Rothman-Denes {\it et al.}{} and Levens {\it et al.}{} (unpublished
results) have used {\it linear} (open) 2300~nm templates
including a T7 RNA polymerase  promoter near the center. Transcription from
this promoter by T7
RNA polymerase generates torsional stress.    Rothman-Denes {\it et al.}{} used
the
activity of a bacteriophage N4 early promoter as a stress reporter.  This
promoter is inactive in its unstressed state and activated through
cruciform extrusion at a superhelical density $\sigma\crit=-0.03$
\cite{dai97a},
corresponding to a torsional stress of $\tau\crit\approx7\eem{14}\,\UT$,
consistent
with the estimate given above\footnote{We estimate that
about 30\% of the superhelical density goes into twisting the
double helix (and the rest into the mean writhe) \cite{volo92a}. Multiplying
$0.3\sigma\crit$ by the microscopic twist stiffness
$C\kbt\approx4.5\eem{19}\,$erg~cm \cite{moro98a} and the relaxed Link
density
$2\pi/(10.5\,{\rm bp}\cdot0.34\,{\rm nm/bp})$ gives the above estimate
for $\tau\crit$. Direct physical manipulation on stretched DNA gives
similar results \cite{stri96a,stri98a}.}. Levens {\it et al.}{} instead used
an element of the human {\it c-myc} gene, which interacts with single-stranded
DNA
binding proteins, and measured unwinding using potassium
permanganate, which
reacts with single-stranded tracts. The results of both sets of experiments
suggest that  structural transitions are induced by T7 RNA polymerase
transcription.    Thus it again appears that {\sl
transcription of linear DNA
can create torsional stress several thousand times greater than
that predicted by the classical formula~(\ref{naive friction}).}

\paragraph{In vivo}
As mentioned above, {\it in vivo} experiments are harder to interpret,
but nevertheless we mention a few illustrative results to show the
very general character of the frictional-drag paradox.

Rahmouni and Wells used a circular 6.3~kb plasmid, reporting its
torsional stress via the B-Z structural transition
\cite{rahm89a,rahm92a}. They concluded that ``the diffusion of
supercoils must be slower than was originally predicted [in
\cite{liu87a}]''.

Lilley and collaborators have carried out an extensive series of
experiments reviewed in \cite{lill96a}. Their conclusion that an ``as
yet unidentified topological barrier should exist'' may point to the
same surprisingly large rotational drag argued for in the {\it in
vitro} experiments above. In later work they also found that the
transcribing polymerase need not be physically anchored, reinforcing
the argument in point {\it (a)} of section~\ref{ssgeneral} above
\cite{chen98a,chen99a}.

Turning finally to experiments in eukaryotes, we mention only
two experiments of Dunaway and coworkers. Dunaway and
Ostrander sought to eliminate any anchoring of their DNA template by
injecting linear DNA with no  subsequences known to associate with the
nuclear architecture into {\it Xenopus} oocytes \cite{duna93a}. They
injected an exogenous (bacterial) polymerase into their oocytes and
ensured that its promoter was the only spontaneously-transcribing
promoter on their template. They also used {\it linear} templates,
reducing the likelihood of any entanglement effects. Using 3.6--4.5~kb
templates with a ribosomal
RNA promoter to report torsional stress, they concluded  that  ``localized,
transient domains of supercoiling'' could occur in open DNA, trapping
significant torsional stress. Similarly, later work by Krebs and
Dunaway  concluded that
``The viscous drag against a large DNA  molecule is apparently
sufficient to prevent transcription-generated supercoils from
diffusing rapidly off the end of the DNA, so DNA length creates a
topological domain'' \cite{kreb96a}. Once again this conclusion is remarkable,
in
that it contravenes the estimates in Sect.~\ref{sintro} above.

\section{Physical picture\label{spp}}
As described in Sect.~\ref{sintro}, the surprising physical aspect of the
experiment is the buildup of torsional stress in the DNA, when
nothing seems to prevent the molecule from spinning almost freely
in place.
Apparently the simple physical model of a uniform elastic rod in a
viscous fluid has left out some crucial effect. One may at this point
be tempted to abandon simple physical models altogether, pointing to
the many specific biochemical features of real DNA which they
omit. But the elastic rod model successfully describes many detailed
features of DNA stretching and fluorescence-depolarization
experiments, including effects of torsional stress ({\it e.g.}
\cite{moro97a,moro98a}). Moreover, the
surprising observed behavior is generic and robust, not specific to a
particular
situation, suggesting that the
model needs only some simple new ingredient in order to
capture the observed behavior.

In this section we argue that augmenting the elastic rod model by
including the {\it natural bends} in the DNA duplex dramatically
changes the transport of torsional stress. The strength of these bends
has been independently measured; it is not a new free parameter. Their
effect on the {\it equilibrium} properties of DNA coils has long been
recognized. In this section and the
next we instead study their effects {\it far} from equilibrium.

\subsection{Need for spin-locking\label{ssnfsl}}
Imagine a given segment of an elastic rod
(modeling a twist-storing polymer
such as DNA) as contained in a black box with only the two ends of the rod
accessible. Cranking one end about its axis amounts to {\sl injecting
a conserved quantity}, ``linking number'' (or \lk), into the
rod.\footnote{Strictly speaking \lk{} is well defined only for a
closed loop. Nevertheless, the {\it change} in \lk{} in an open
segment with fixed end is well defined, and must vanish, whatever
happens inside the black box. Rotating one end about its
axis thus injects a conserved quantity.} We can
schematically think of linking number as taking one of five pathways
away from the cranking site:
\penumerate
\item \lk{} can be elastically stored as {\it twist} in the rod: the rod
segment can rotate about its axis by an amount which depends on
position along the rod;
\item \lk{} can be elastically stored as {\it writhe}: the rod can begin
to supercoil;
\item \lk{} can be transported by {\it spinning} (plumber's-snake)
motion, emerging at the far end with no net change in the rod state;
\item \lk{} can be transported by {\it rigid rotation} (crankshaft
motion) of the whole segment about some axis;
\item \lk{} can be {\it lost} via the action of topoisomerase.
\end{enumerate}
We are interested in steady-state transport, in the
absence of topoisomerase, and so we consider only the competition
between pathways \#3 and \#4.

This picture allows a more precise summary of the paradox reviewed in
Sects.~\ref{sintro}--\ref{sexperiments} above. The steady transport of
injected \lk{} will meet with resistance in the form of effective
frictional constants $\mu\spin$ for spinning and $\mu\rigid$ for rigid
rotation, and hence a total frictional constant
$\mu\tot=({\mu\spin}\inv+{\mu\rigid}\inv)\inv$. But we have seen that
experimentally $\mu\tot$ is much larger than the theoretically
expected value of $\mu\spin$. No matter how large $\mu\rigid$ may be,
it cannot resolve this paradox. In particular the well-known
coupling between torsional stress and writhing motion
(see \eg{}~\cite{shi94a,coll96a} and references therein) is of no help,
since the problem is precisely that there is little torsional stress.

What is needed is a way to {\it shut down} pathway \#3, \ie{} to {\it
lock} the spin degree of freedom, at least partially.

The fact that a uniform rod is never actually straight on length
scales beyond its bend-persistence length $\Asym$ does not help,
either.\footnote{Even in the
absence of thermal motion, a naturally-straight rod will bend when
cranked fast enough, executing a hybrid of rigid rotation and spinning
\ifx\ifPNAS\figtest (C. W. Wolgemuth, T. R. Powers, and R. E. Goldstein,
unpublished
results). \else\cite{wolg9?}. \fi Wolgemuth {\it et al.} found, however, that
for the
parameters of interest to us here the \lk{} transport is dominated by
spinning, exactly as argued above. \ifx\ifPNAS\figtest\else The ease of
spinning relative to translation of the rod
through the fluid can alternately be understood from the point of
view of Brownian fluctuations: a thin, axially-symmetric object
receives random thermal kicks from the surrounding fluid, but
these deliver very little torque due to the small rod radius
$R$. By the general relation between diffusion and friction
\cite{berg93a}, we again obtain a  rotational friction
constant suppressed by powers of $R$, consistent with
formula~(\ref{naive friction}). \fi
}  Spinning creates no long-range
hydrodynamic interaction, since the fluid velocity field falls
off on the scale of the rod diameter $R=1\,$nm
\cite{lamb45a}.  Since $\Asym$ is much larger than  $R$, the straight rod
approximation is
adequate \cite{schu97a}. Certainly the spinning in place of a
thermally-bent but naturally-straight rod requires
continuous flexing of the rod, as the direction of curvature
rotates in the material frame of the rod, but
the elastic cost of a bend in a cylindrical rod depends only on
the magnitude, not the direction, of the curvature, and this does
not change: such a rod has no energetic barrier to spinning.

To summarize, the na\"\i{}ve equation (\ref{naive friction}) will be
accurate, and torsional stresses will be small, unless some
sort of locking mechanism {\it inhibits free spinning} of
linear DNA in solution. To find such a mechanism, we must now introduce
some new element of realism into our description of DNA.

\subsection{Natural bends\label{ssbends}}
As mentioned in Sect.~\ref{sintro}, the key ingredient missing so far
from our model is the natural curvature of the DNA duplex. Immense
effort has been focused on predicting the precise conformation of
a DNA tract given its basepair sequence, using molecular modeling,
oligomer crystallography, and NMR, among other
techniques. Fortunately, for our problem it suffices to characterize
the {\it average} effect of curvature over hundreds of basepairs. For
such purposes a very simple phenomenological approach suffices.

Natural DNA is a stack of similar but nonidentical subunits,
arranged in an order which is fixed but random for our purposes.
It is crucial that even though these bends
are random, their effects do not average to zero on length scales much
longer than one
base-pair. Instead, the minimum-energy conformation of such a stack
may be regarded as a distorted helix whose backbone follows a
random walk, with a {\it structural persistence length} $\Psym$. Note
that $\Psym$ is a purely geometrical parameter, having nothing to do
with the mechanical bend stiffness $\kasym$ of DNA nor the thermal
energy $\kbt$. Instead $\Psym$ reflects the {\it information content}
in a piece of DNA.

Just as in the straight case, bent (natural) DNA can also be deformed
{\it away}
from its minimum-energy state at some enthalpic cost characterized by
a bend stiffness $\kasym$, with
units energy$\cdot$length. Since fluctuations are controlled by the
thermal energy $\kbt$, we define the {\it bend
length} $\Asym=\kasym/\kbt$. The combined
effect of thermal and natural bends then makes DNA a random coil with
total persistence length\footnote{Some authors call $P$ the ``static
persistence length'' and $A$ the ``dynamic persistence
length''. Schellman and Harvey verified Trifonov {\it et al.}'s heuristic
derivation of this formula within a number of detailed models
\cite{sche95a}. Since $\Psym\inv<\Asym\inv$ we can regard the
bend disorder as smaller than the thermal disorder. In this case
Trifonov's formula  also gives the effective
persistence length measured by fitting DNA stretching experiments to the
na\"\i{}ve worm-like chain model \cite{nels98b}.} $\Atsym=\bigl(
\Asym\inv+\Psym\inv\bigr)\inv$
\cite{trif87a}.
Under physiological conditions $\Atsym$ has the familiar value of
50~nm. Experiments on artificial, naturally-straight DNA make it
possible to determine $\Asym$ and $\Psym$ separately, yielding
$\Asym\approx80\,$nm and $\Psym\approx130\,$nm
\cite{bedn95a}.\footnote{
Though Bednar {\it et al.} did not estimate
the uncertainty in their determination of $\Psym$ \cite{bedn95a}, it
may well be
large. They note, however, that their direct
experimental determination agrees with the model-dependent prediction
of Bolshoy {\it et al.} \cite{bols91a}.}

\subsection{Hybrid motion\label{sshm}}
We wish to explore the consequences of the natural bends introduced in
the previous subsection for the transport of torsional stress in
DNA. Before doing any calculations, it is worthwhile to formulate some
intuitive expectations, based on four increasingly realistic cartoons for the
steady-state motion of a
cranked DNA segment of contour length $\ell$ (\tfig{ffour}{\it
a--d}\/).

\ifigure{ffour}{Four increasingly realistic models of cranked DNA
motion.
{\it a)}~Straight, rigid rod, assumed in the derivation of the na\"\i
ve formula eqn.~(\ref{naive friction}).
{\it b)}~Naturally-straight but thermally-bent rod.
{\it c)}~Naturally-bent, rigid rod.
{\it d)}~Hybrid motion of a naturally-bent, semiflexible rod. The rod
rotates rigidly on length scales shorter than $L_C$ while flexing on
scales longer than $L_C$.
}{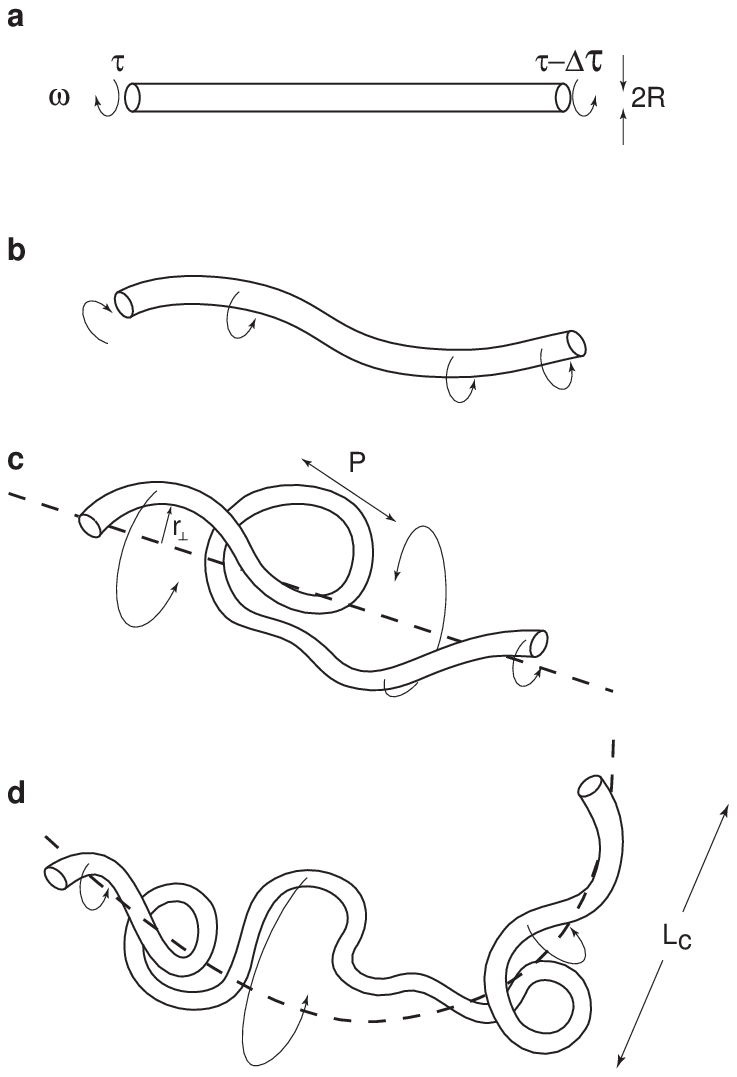}{3.6} %3.6

As noted in Sect.~\ref{sintro}, a straight, rigid segment
(\tfig{ffour}{\it a}\/) would encounter a torsional drag per unit length
$\mu\spin\omega$, or a net drop in torsional stress between the ends
of $\mu\spin\omega\ell$, with friction constant $\mu\spin$ given below
eqn~(\ref{naive friction}). We argued in Sect.~\ref{ssnfsl} that the
case of a naturally-straight but semiflexible segment is similar
(\tfig{ffour}{\it b}\/).

Matters change considerably when we introduce natural bends. If the
rod were perfectly rigid (\tfig{ffour}{\it c}\/), it would have to execute
crankshaft motion; individual rod elements would then drag sideways through the
fluid. We
will see below that as $\ell$ increases, the corresponding drag per unit
length would increase without bound. On long enough scales, then, we may
expect that any realistic molecule  cannot be regarded as infinitely
stiff.

At the other extreme, we could imagine the naturally-bent rod
spinning in place. This however would mean that every joint
periodically bends {\it oppositely} to its preferred conformation. The
corresponding elastic energy cost creates a barrier to this motion.

We will argue that in fact a real, semiflexible heteropolymer chooses
a compromise between these extremes of motion, selecting a {\sl
crossover scale} $L_C$ and executing a hybrid motion (\tfig{ffour}{\it
d}\/). On length scales shorter than $L_C$ this motion is nearly
rigid, since as just argued an activation barrier resists flexing. On
longer length scales the motion must cross over to spinning,
since as just argued rigid (crankshaft) motion meets a large viscous
drag on long scales.

We must now justify these intuitive ideas and obtain a numerical
estimate for the crucial crossover scale $L_C$. Since $L_C$ will turn
out to be significantly longer than the basepair step size, we will
conclude that the spinning  (plumber's-snake) motion is effectively locked, as
we argued
was necessary in Sect.~\ref{ssnfsl}.

\section{Scaling Analysis\label{scalcs}}
We must now justify and quantify the expectations sketched in
Sect.~\ref{spp}.

\subsection{Spin-locking\label{sssl}}
Consider first the hypothetical case of a perfectly rigid,
naturally-bent rod (\tfig{ffour}{\it
c}\/). The viscous force per length $f$ on a
straight rod much longer than its radius $R$, dragged sideways through
a viscous medium, is \cite{garc81a}
\begin{equation}
f \equiv \mu\drag v={4\pi\eta\over0.8+\ln (X /2R)}\,v
\label{trans drag}\end{equation}
where $v$ is the speed and $X$ is the rod length. Our polymer is of course
{\it not} straight on length scales beyond its structural persistence
length $\Psym$, so we substitute
$\Psym$ for the long-scale cutoff $X$ in eqn.~(\ref{trans drag}). Since the
dependence on $X$ is weak this is
a reasonable approximation.\footnote{A rod pulled at some angle other than
$90^\circ$ to its tangent will have a drag given by eqn.~(\ref{trans
drag}) with a slightly different prefactor; we will neglect this
difference and use eqn.~(\ref{trans drag}) in all cases.}
Taking $\Psym=130\,$nm and $R=1\,$nm then
gives $\mu\drag\approx2.5\eem2\,\UV$.

Suppose we crank a rod segment of arc length $\ell$, which then rotates
rigidly about an axis.  Each element of the rod then moves through
fluid at a speed $v=r_\perp\omega$, where $r_\perp$ is the distance
from the rod element to the rotation axis (\tfig{ffour}{\it
c}\/). Multiplying the moment arm $r_\perp$
times the drag force, eqn.~(\ref{trans drag}), and integrating over
the curve yields the torque drop $\Delta\tau=\mu\drag\omega
\ell\langle r_\perp^{\ 2}
\rangle$ across the segment. Here $\langle{ r_\perp}^2\rangle$ is the
average of ${ r_\perp}^2$ along the rod segment.

Each rod segment of course has a different sequence, and hence a
different preferred shape. Each segment will therefore have a
different value of $\langle{ r_\perp}^2\rangle$. Fortunately, we are
interested in the sum of the torque drops across many segments, each
with a different, random, sequence. Thus we may replace $\langle{
r_\perp}^2\rangle$ by its {\it ensemble average} over sequences, which
we will call  $\llangle{r_\perp}^2\rrangle$. This average has a simple
form: eqn.~7.31 of
Ref.~\cite{yamakawabook} gives $\llangle{{\bf r}_\perp}^2\rrangle=\ell
P/9$, and hence
\begin{equation}\Delta\tau=\mu\drag\omega \ell^2P/9
\label{edrag}\end{equation}
In the language of Sect.~\ref{ssnfsl}, we have just estimated the drag
torque $\mu\rigid\omega\ell$, finding $\mu\rigid\approx\mu\drag\ell P/9$.
Indeed we see that the drag per unit length grows with $\ell$, as
suggested in Sect.~\ref{sshm} above.  Formula (\ref{edrag}) is valid when
the segment length $\ell$ is longer  than $\Psym$, an assumption
whose self-consistency we will check below.

We can now relax the artificial assumption of a perfectly rigid rod
and thus
pass from \tfig{ffour}{\it c}\/ to the more realistic \tfig{ffour}{\it d}.
Suppose that a long polymer has been subdivided into segments of
length $\ell$, each approximately executing rigid rotation about a
different axis. The axes will all be different, since we are assuming
that $\ell$ is longer than the structural persistence length $\Psym$.
To join these segments smoothly as they rotate, each segment therefore
needs to flex. On average, each segment must periodically bend one end
relative to the other by about $90^\circ$. The least costly conformational
change which accomplishes this is to spread the bending strain uniformly along
the entire segment length $\ell$; we can then estimate the elastic
bending-energy cost as \footnote{Natural DNA can have localized
regions of reduced bend stiffness. These flexible tracts will not
significantly affect this estimate unless they are spaced more closely
than the length scale $L_C$ found below.}
\ifx\ifPNAS\figtest${\ell\kasym\over2}\Bigl({\pi\over2\ell}\Bigr)^2$. \else
\begin{equation}
{\ell\kasym\over2}\Bigl({\pi\over2\ell}\Bigr)^2
\label{ebarrier}\end{equation}\fi
This energy barrier becomes small for large $\ell$, just the opposite
trend to that of eqn.~(\ref{edrag}). The physical reason for this
behavior is that we do not insist on ironing out every small kink
in the rotating rod's shape; the rod segment can satisfy the
imposed conditions on its ends by deforming only a
fraction of its many intrinsic bends.

The bending energy needed to crank the segment through an angle
$\theta$ is  roughly \ifx\ifPNAS\figtest the above expression \else
eqn.~(\ref{ebarrier}) \fi times
$\half(1-\cos\theta)$; the torque needed to
increase $\theta$ is then the derivative of this formula,
$\half\sin\theta$. Thus the driving torque needed to overcome the
bending-energy barrier
 turn through a complete revolution is just one half
of \ifx\ifPNAS\figtest the above expression. \else
eqn.~(\ref{ebarrier}). \fi The crossover length $L_C$ is then the value
of $\ell$ at which  the viscous torque drop,
eqn.~(\ref{edrag}),  just balances this critical value:
%\ifx\ifPNAS\figtest $\mu\drag\omega{L_C}^2P/9={\kbt}{A}{\pi^2}/16L_C$.\else
\begin{equation}
\mu\drag\omega{L_C}^2P/9={\kbt\over2}{A\over L_C}{\pi^2\over8}
\label{ecrossover}\end{equation}
Substituting the numerical values we find $L_C\approx450\,$nm for T7
RNAP, and slightly larger for other, slower, polymerases.

Our crossover length has indeed proven to be longer than the
structural persistence length $\Psym$, so the assumption
$\ell> \Psym$ made above
is self-consistent.
Indeed, $L_C$ has proven to be about 1.4~kbp. In
our illustrative example of a 7~kbp DNA construct cranked at the
midpoint, we see that intrinsic bends shut
down spinning motion almost completely: {\sl the na\"\i{}ve model of
Sect.~\ref{sintro} does not describe the true motion at all}. We must
now see what this implies for the overall torsional stress on the
construct.

\subsection{Hydrodynamic interactions\label{sshi}}
In contrast to spinning in
place, dragging a thin rod sideways sets up a long-range flow
field. Now that we know that spinning is effectively forbidden, we
must therefore study the possibility of long-range hydrodynamic interactions
between rod segments.

The theory of polymer dynamics tells us that a
short random coil dragged through fluid can be viewed as a set of
thin-rod elements
moving independently in a motionless background (the
``free-draining'' case), but a long
coil instead moves as a {\it solid spherical object}, due to hydrodynamic
interactions~\cite{hiem84a}. The crossover between these two
regimes is controlled by the dimensionless parameter
$Q\equiv\sqrt{{L\over\Asym\tot}}{\mu\drag\over\eta}$. Free
draining corresponds to the case $Q\ll1$.
For our illustrative example of a coil of length $L=2300\,$nm and
total persistence length $\Asym\tot=50\,$nm, we
get $Q=17$,  interactions are important, and the coil moves as a
solid sphere.

The viscous drag torque on such a coil is $\tau=\mu_{{\rm
coil}}\omega L$, where $\mu_{{\rm
coil}}={4\over9}\sqrt{3\pi^3LA^3}\cdot1.26\eta$ (see \S31 of
\cite{yamakawabook}). Dividing this torque equally between the
upstream and downstream halves of the construct, we find the estimated
torsional stress on either side of the cranking point to be
$\omega\cdot1.0\eem{15}\,$dyn~cm~sec. Taking
$\omega=60\,$radian/sec then gives a torsional stress of
$6\eem{14}\,\UT$, comparable to the
value quoted in Sect.~\ref{sexperiments}
as necessary to induce structural transitions and {\sl about seven
thousand times greater than
the na\"\i ve estimate} given below eqn.~(\ref{naive friction}).

\subsection{Relation to prior theoretical work\label{ssotherwork}}

The viewpoint taken in this paper can be regarded as a synthesis of
two established threads.

\paragraph{Fluid-mechanics work} One of these threads studies the
deterministic dynamics of
externally driven ({\it i.e.} far from equilibrium) rods in a viscous
environment. For example, Garcia de la Torre and Bloomfield studied
the effects of a {\it single, permanent, large-angle} bend on viscous
drag \cite{garc81a}, obtaining precise versions of some of the
formul\ae{} given above. Individual large-angle bends caused by
DNA-binding factors may well be present {\it in vivo}, but our point
here is that a {\it statistical distribution} of {\it small,
finite-stiffness} bends still leads to dramatic effects.

Several authors have studied the interplay between shape and twist in
the dynamics of {\it naturally straight, flexible} rods in a viscous
medium \ifx\ifPNAS\figtest \cite{gold95a,kami98a,gold98a}, \else
\cite{gold95a,kami98a,gold98a,wolg9?}, \fi again obtaining precise
formul\ae{} for situations simpler than that studied here. It would
be very interesting to incorporate intrinsic bends into their
formalism.

Finally, Marko has proposed that the {\it impulsive} (jumpy) action of
RNA polymerase can lead to {\it transient} torsional stresses greater
than predicted by the na\"\i ve formula, eqn.~(\ref{naive friction})
\cite{mark98a}. The range of this enhancement, however, depends on the
time scale of each step and may be too short to explain the observed
phenomena. Experimental measurement of this time scale will be needed
to assess this proposed mechanism.

\paragraph{Simulation work} A second thread is the extensively studied problem
of the
{\it equilibrium fluctuations} of a polymer, particularly
the diffusive torsional motion of DNA as measured in fluorescence experiments.
Most of this work used Monte Carlo or Brownian dynamics numerical simulation
techniques; most did not introduce long-range hydrodynamic
interactions as we did in Sect.~\ref{sshi} above.

Fujimoto and Schurr noted that fitting experimental fluorescence
polarization anisotropy data to a model of intrinsically-straight DNA
yielded an effective hydrodynamic radius which increased with
increasing segment length \cite{fuji95a}. They suggested the
possibility that this effect could be caused by permanent or
long-lived bends in DNA.

Collini {\it et al.}  took up the same problem \cite{coll96a},
explicitly introducing intrinsic bends. Their physical model, however,
was the crankshaft motion of a perfectly rigid,
zig-zag shape. The zig-zag shape introduces structure on {\it one} length
scale.
A major point of the scaling analysis in Sect.~\ref{scalcs} above, however,
was that the minimum-energy conformation of natural DNA is  actually a
random coil, and random walks have structure on {\it all} length
scales. A second key point of our analysis
was that DNA is {\it not} infinitely stiff, leading to the crossover
phenomenon found in Sect.~\ref{sssl}.

Schurr {\it et al.} distinguished between ``phase-locked bends'',
equivalent to the natural bends in the present work, and
``non-phase-locked bends'' including the thermal bends of the present
work.\footnote{Another example of a non-phase-locked bend could be a
universal joint: a bend maintaining fixed polar angle but
free to swivel in the azimuthal direction. Schurr {\it et al.} also
distinguish between slowly- and rapidly-relaxing bends.
The present work assumes that the large
external applied torsional stress (absent in the equilibrium situation
studied in \cite{schu97a}) suffices to overcome any kinetic barriers
to elastic deformation of the DNA duplex.} They verified using Monte Carlo
simulation that in the absence
of natural bends, the torsional drag on a thermally-bent rod is the
same as that for a straight rod, as argued physically in
Sect.~\ref{ssnfsl} above. Schurr {\it et al.} went on to anticipate
the hybrid motion studied in the present work, proposing that ``beyond
some length the degree of global phase locking should decrease, as the
motion approaches that of a wobbly eccentric speedometer cable, and
the effective hydrodynamic radius should reach a plateau value,
which is possibly 1.2~nm. The available evidence indicates that this
radius is independent of length for $L>60\,$nm'' \cite{schu97a}. The
authors did not, however, present a model incorporating
random natural-bend disorder.

The present work predicts instead that the response of DNA to external
cranking is controlled by an effective drag constant that does not
saturate until $L>L_C$. The crossover scale $L_C$ depends on the
transcription rate via eqn.~(\ref{ecrossover}) and is typically
hundreds of nanometers; the saturation value of the effective
hydrodynamic radius is then
much greater than  1.2~nm. The driven situation of interest
here is not, however, the same as the equlibrium situation studied in
\cite{schu97a}.

Finally,  A.~Maggs has independently shown that
in a naturally-straight, thermally-bent rod twist relaxation follows
the same diffusive law as in a rigid straight rod, out to
extremely long scales (over 2~kbp) (A. Maggs, unpublished
results). Beyond this scale
Maggs found that pathway \#2 in Sect.~\ref{ssnfsl} above begins to
affect twist relaxation, leading to an interesting new scaling
relation.
\section{Conclusion}
The analysis of this paper rests upon a surprising fact from
slender-body viscous hydrodynamics.  The drag torque for spinning a
thin rod behaves reasonably as one decreases the rod radius $R$: it is
proportional to
$R^2$. In surprising contrast, the drag force for pulling such a rod
{\it sideways} is practically independent of $R$ (eqn.~(\ref{trans drag})
above). The only length scale available to set the rotational drag for
rigid crankshaft motion is then the radius of curvature of the
rod. But a randomly-bent rod has structure on every length scale, and
so the drag torque per length increases without bound for longer
segments until the crossover condition, eqn.~(\ref{ecrossover}), is
met. Since the crossover scale $L_C$ proves to be long, cranked DNA is
effectively spin-locked on scales shorter than at least 1~kbp. This
observation explains why the na\"\i ve formula, eqn.~(\ref{naive
friction}), is inapplicable, eliminnating the paradox described in
Sect.~\ref{sintro}.

The transport of torsional stress may enter in many cell
processes. While this paper has stressed its possible role in gene
regulation, torsional stress has recently been assigned a role in the
disassembly of nucleosomes in front of an advancing polymerase complex
({\it e.g.} \cite{jack94a,gall95a}), in
chromatin remodeling ({\it e.g.} \cite{lee91a}), and in the action of
enzymes on DNA ({\it e.g.} \cite{vill84a}). The
ideas of this paper may be relevant to these problems too, though of
course in eukaryotes the phenomenon described here may be
preempted by the effects of higher-order chromatin structure. Direct
manipulation of single DNA molecules sometimes involves cranking as
well ({\it e.g.} \cite{esse97a}).

The simple scaling analysis used in this paper makes some testable
predictions.
The key claim has been that intrinsic bends can have
a huge effect on the transport of torsional stress along DNA. For
example, synthetic DNA engineered to be less bent
than natural sequences~\cite{bedn95a} will have longer crossover scale
$L_C$ (eqn.~(\ref{ecrossover})), and
hence should support less torsional stress for a given length. Shortening
a linear template below $L_C$ should also sharply reduce the overall drag
coefficient. More generally, none of the experimental papers cited
earlier made quantitative estimates of the effective torsional
friction constant needed to explain their results. One could imagine
an {\it in vitro} experiment using  local stress reporters ({\it e.g.}
the B-Z structural transition), inserted at various positions, to get
the full
torsional stress profile, in space and time, as function of transcription
rate. Even a limited subset of this quantitative information would
yield insight into the mechanisms of torsional stress transport.

\section*{Acknowledgments}
I wish to thank S.~Block, N. Dan, P. Dr\"oge, M. Dunaway, R.~E.~Goldstein,
J.~Marko,
T.~R.~Powers,
J.~M. Schurr, and
C.~Wiggins,
for valuable discussions, and particularly
D. Levens, A. Maggs, and L.~B. Rothman-Denes for describing their
unpublished work.
This work was supported in part by  NSF grant DMR98-07156.

\newpage
\ifx\ifPNAS\figtest
\bibliographystyle{pnas}   %submit
\else
\bibliographystyle{unsrt}  %preprint
\fi
%\bibliography{dna,torque}

\end{document}